\pdfoutput=1
\documentclass[a4paper]{article}
\usepackage{graphicx}
\usepackage{fullpage}
\newcommand{\m}{{\check m}}
\newcommand{\N}{{\tilde N}}
\newcommand{\p}{{\tilde p}}
\newcommand{\ve}{\mathbf}

\newcommand{\up}{\uparrow}
\newcommand{\down}{\downarrow}
\newcommand{\gs}{{\scriptstyle {}^{>}_{\sim}}}
\begin{document}
\title{
Threshold photoproduction of neutral pions off protons in nuclear
model with explicit mesons
	}
\author{
	D.V.~Fedorov\footnote{fedorov@phys.au.dk}
	~and M.~Mikkelsen \\
	{\it Aarhus University, Aarhus, Denmark}
}
\date{}
\maketitle

   \begin{abstract}
We apply the nuclear model with explicit mesons to photoproduction of
neutral pions off protons at the threshold.  In this model the nucleons do
not interact with each other via a potential but rather emit and absorb
mesons that are treated explicitly on equal footing with the nucleons.
We calculate the total cross section of the reaction for energies close to
threshold and compare the calculations with available experimental data.
We show that the model is able to reproduce the experimental data and
determine the range of the parameters where the model is compatible with
the experiment.
   \end{abstract}

   \section{Introduction}

In~\cite{fedorov-sigma} a nuclear model with explicit mesons has
been proposed where the nucleons do not interact with each other via
a potential but rather emit and absorb mesons.  The mesons are treated
explicitly on equal footing with the nucleons.  It has been shown---using
the sigma-meson in a proof of concept calculation---that the model is
able to reproduce the bound state of a neutron and a proton, the deuteron.

One possible advantage of this model over the conventional interaction
models---where the particles interact directly via a phenomenological
potential---is the reduced number of parameters.  There are basically
only two parameters per meson: the range and the strength of the
meson-nucleon coupling.  For example, in this model the two parameters
for the pion-nucleon coupling determine all the forces mitigated by the
pions: the central, tensor, and the three-body force (as the two-pion
exchange force).  Moreover these two parameters also determine the
low-energy physics with pions, in particular the photoproduction of pions
off nucleons.  Should the model turn out to actually work in practice it
must be able---with the two parameters for pions plus two parameters for
sigma-mesons---to describe the deuteron, the low-energy nucleon-nucleon
scattering, the pion-nucleon scattering, the pion photoproduction,
and---with the resulting tree-body force---also the triton.

In this investigation we are going to apply the model to neutral pion
photoproduction off protons where there exist precise experimental
data~\cite{bergstrom,fuchs,schmidt}.  We shall calculate the total cross
section for this reaction and then compare it with the experimental data
with a view to determine the model's pion parameters that are compatible
with the experimental data, if there are any.

We first introduce the model with one possible pion-nucleon coupling
operator that conserves isospin, angular momentum, and parity; we
then derive the equations describing the dressing of bare protons with
pions in one-pion approximation; derive the formulae for the neutral
pion photoproduction cross section off the dressed proton; calculate the
total cross-section, compare it with the experimental data, and determine
the range of the parameters of the model where it is compatible with
the experiment.

\section{Nuclear interaction model with explicit pions}

In the model with explicit pions the nucleons do not interact directly
via a potential but instead emit and absorb pions via a suitable
pion-nucleon coupling operator.  In the absence of any extra energy
the pion emitted by a nucleon is virtual: it cannot leave the nucleon
as it finds itself in a classically forbidden region under a potential
barrier about the size of the mass of the pion. A bare nucleon surrounded
by virtual pions is called a physical or dressed nucleon.  A dressed
nucleon exists in a superposition of states with different number of
pions.	The corresponding multi-component wave-function of the dressed
nucleon, $\Psi_N$, can be written as
	\begin{equation}
\Psi_N=\left( \begin{array}{c} \psi_\N \\ \psi_{\N\pi} \\ \psi_{\N\pi\pi}
\\ \vdots \\ \end{array} \right)
	\,,\end{equation}
where $\psi_\N$ is the wave-function of the bare nucleon, $\psi_{\N\pi}$
is the wave-function of the bare nucleon plus one pion, $\psi_{\N\pi\pi}$
is the wave-function for the bare nucleon plus two pions and so on.
The Hamiltonian that acts on this multi-component wave-function is
given as
	\begin{equation}
H=\left(
\begin{array}{cccc}
\m_\N +K_\N      & W^\dagger               & 0 & \dots \\
W & \m_\N+K_\N+\m_\pi+K_\pi+V_C  & W^\dagger & \dots \\
0 & W &  \m_\N+K_\N+2\m_\pi+K_{\pi_1}+K_{\pi_2}+V_C &  \dots \\
\vdots    & \vdots          & \vdots   & \ddots 
\end{array}
\right) \,,
	\end{equation}
where $K_\N$ is the kinetic energy of the bare nucleon, $K_\pi$
is the kinetic energy of the pion,
$\m_\N\equiv m_\N c^2$ and $\m_\pi\equiv m_\pi c^2$
are the masses of the bare nucleon and the pion
in units of energy,
$V_C$ is the Coulomb interaction between the charged particles, if any,
and $W$ ($W^\dagger$) is the operator that generates (annihilates) a pion.
Assuming that nuclear interaction conserves
isospin, angular momentum, and parity,
one possible operator that generates a pion---a scalar
isovector with negative parity---can be written
as~(cf.~\cite{siemens-jensen})  
	\begin{equation}
W(r) =
(\vec\tau\vec\pi)
(\vec\sigma\vec r)f(r) \;,
       \end{equation}
where $\vec\tau$ is the isovector of Pauli matrices that act in the
isospin space of the nucleon, $\vec\pi$ is the isovector of pions,
$\vec\sigma$ is the vector of Pauli matrices that act in the spin space
of the nucleon, $\vec r$ is the relative coordinate between the nucleon
and the pion, and $f(r)$ is a phenomenological (short-range) form-factor.
The isospin factor $\vec\tau\vec\pi$ is given as
	\begin{equation}
\vec\tau\vec\pi=
\tau_0\pi^0+\sqrt{2}\tau_-\pi^++\sqrt{2}\tau_+\pi^- \;,
	\end{equation}
where $\pi^0$, $\pi^+$, and $\pi^-$ are the physical pions and where
the $\tau$-matrices are given as
	\begin{equation}
\tau_0=\left(
\begin{tabular}{cc}
1 & 0 \\
0 & -1
\end{tabular}
\right)\,,\;
\tau_-=\left(
\begin{tabular}{cc}
0 & 0 \\
1 & 0
\end{tabular}
\right)\,,\;
\tau_+=\left(
\begin{tabular}{cc}
0 & 1 \\
0 & 0
\end{tabular}
\right)\,.
	\end{equation}
The $W$ operator adds an extra pion to a state of nucleons and pions
and the $W^\dagger$ operator annihilates a pion\footnote{The action
of $W^\dagger$ includes integration $\int_V d^3r$ to get rid of the
coordinate of the annihilated pion.}.

The ground state of the corresponding Schr\"odinger equation,
	\begin{equation}
H\Psi_N=E\Psi_N \,,
	\end{equation}
is the physical (dressed) nucleon -- a bare nucleon surrounded by a
cloud of under-the-barrier (virtual) pions. The ground state energy
in the rest frame of the nucleon gives the mass of the physical nucleon.

\section{Dressing of the proton in one-pion approximation}

Generation of a pion requires more than $\m_\pi\approx $140~MeV of
extra energy. In the absence of the extra energy the pion generated by
a bare proton finds itself in a classically forbidden region
under the potential barrier of those 140~MeV. Its wave-function falls
off asymptotically as $e^{-r/\lambda}$ where $\lambda\approx\hbar
c/\m_\pi\approx$1.4~fm. The pion is virtual as it cannot leave the proton.

Given the amount of energy in excess of the pion mass (plus recoil)---from
an incoming photon, for example---the virtual pion can be knocked off
the proton and become a physical pion. That would be what is called a
pion photoproduction reaction.

Each additional virtual pion adds yet another 140~MeV to the potential
barrier under which the corresponding system of particles resides.
Therefore one might assume that the first virtual pion---which is under
the smallest barrier---should account for the largest contribution to
the dressing of the proton.  That is the one-pion approximation to which
we are going to restrict ourselves in this investigation.
This assumption must be of course checked by direct calculations. However
adding a second pion is a much bigger task which will hopefully be dealt
with in a separate investigation.

In the one-pion approximation
the physical proton is a superposition of two coupled components: the
bare-proton component and the bare-nucleon plus one pion component. The
Hamiltonian for the physical proton is then the given as
       \begin{equation}\label{eq:ham2}
H= \left( \begin{array}{cc} K_\p +\m_\N & W^\dagger  \\ W & K_\N+\m_\N
+K_\pi+\m_\pi \end{array} \right)
       \,.\end{equation}
The Hamiltonian acts on a two-component wave-function of the physical
proton,
       \begin{equation}
\Psi_p= \left( \begin{array}{l} \psi_\p (\vec R)\\ \psi_{\N\pi}(\vec
R,\vec r) \end{array} \right) \,,
       \end{equation}
where $\vec R$ is the center-of-mass coordinate and $\vec r$ the relative
coordinate between the pion and the nucleon.  The kinetic energy operators
are given as
	\begin{eqnarray}
K_\p &=& \frac{-\hbar^2}{2m_\N}\frac{\partial^2}{\partial\vec R^2} \,, \\
K_\N + K_\pi &=& \frac{-\hbar^2}{2M_{\N\pi}}\frac{\partial^2}{\partial\vec
R^2} + \frac{-\hbar^2}{2m_{\N\pi}}\frac{\partial^2}{\partial\vec r^2} \,,
	\end{eqnarray}
where $m_\N$, $m_\pi$ are the masses of the bare nucleon and the
pion, $M_{\N\pi}=m_\N+m_\pi$ is the total mass, and $m_{\N\pi}=m_\N
m_\pi/(m_\N+m_\pi)$ is the reduced mass of the bare nucleon and the pion.

In the rest frame of the proton the dependence on $\vec R$ disappears
and $\psi_\p$ can be chosen as
	\begin{equation}
\psi_\p=\frac{p\up}{\sqrt{V}}
	\,,\end{equation}
where $p$ is the isospin state of the proton,
	\begin{equation}
p=\left( \begin{array}{c} 1 \\ 0 \end{array} \right)
	\,,\end{equation}
($\tau_0p=p$), and
$\up$ is the spin state of the proton,
	\begin{equation}
\up=\left( \begin{array}{c} 1 \\ 0 \end{array} \right)
	\,,\end{equation}
($\sigma_0\up=\up$).
The function $\psi_\p$ is normalized to one proton in the volume $V$.

The corresponding two-component Schr\"odinger equation is given as
	\begin{eqnarray}\label{sch-2}
\int_V d^3r (\vec\tau\vec\pi)^\dagger
(\vec\sigma\vec r)^\dagger f(r)\psi_{\N\pi} &=&
\tilde E \frac{p\up}{\sqrt{V}} \,,\nonumber\\
(\vec\tau\vec\pi)
(\vec\sigma\vec r)f(r)\frac{p\up}{\sqrt{V}}
+ \left(
\frac{-\hbar^2}{2m_{\N\pi}}\frac{\partial^2}{\partial\vec r^2} + \m_\pi
\right)\psi_{\N\pi} &=& \tilde E \psi_{\N\pi}
	\,,\end{eqnarray}
where $\tilde E\equiv E-\m_\N$.
The second row of the equation indicates
that $\psi_{\N\pi}$ must have the same spin-isospin structure as
	\begin{equation}
(\vec\tau\vec\pi)(\vec\sigma\vec r)p\up
	\,,\end{equation}
suggesting that one can
search for $\psi_{\N\pi}$ in the form
	\begin{equation}
\psi_{\N\pi}
=(\vec\tau\vec\pi)(\vec\sigma\vec r)
\frac{p\up}{\sqrt{V}}\varphi(r)
	\;,\end{equation}
where $\varphi(r)$ is a scalar radial function with the dimension
length$^{-5/2}$ such that the integral
	\begin{equation}
\int_V d^3R \int_V d^3r |\psi_{\N\pi}|^2
	\end{equation}
is dimensionless.

Using the formulae
	\begin{eqnarray}
(\vec\tau\vec\pi)^\dagger(\vec\tau\vec\pi) &=& 3 \,,\\
(\vec\sigma\vec r)(\vec\sigma\vec r) &=& r^2 \,,\\
\nabla^2(\vec r\varphi(r))
	 &=& \vec r(\varphi''+\frac{4}{r}\varphi') \;,
	\end{eqnarray}
the Schr\"odinger equation~(\ref{sch-2}) can be rewritten as
	\begin{eqnarray}\label{sch-3}
3\int_Vd^3r~r^2f(r)\varphi(r)&=&\tilde E \nonumber \\
-\frac{\hbar^2}{2m_{\N\pi}}(\varphi''+\frac{4}{r}\varphi')+f(r)
&=&(\tilde E-\m_\pi)\varphi
	\,,\end{eqnarray}
where the boundary conditions for $\varphi(r)$ are (assuming that
the formfactor $f(r)$ is finite and short-ranged)
   \begin{equation}\label{bcond}
\left\{
\begin{array}{lcl}
\varphi(r\to 0) &\to& \mathrm{const} \;,\\
\varphi(r\to\infty) &\to&  0 \;.
\end{array}\right.
   \end{equation}
The constant and the ground state energy, $\tilde E_0$, must be found
from the self-consistent solution to the equations where the boundary
conditions are satisfied and where the mass of the bare proton is chosen
such that the ground state energy gives the experimental mass of the
proton, $\m_N=\m_\N+\tilde E_0$.

Notice that
	\begin{equation}
(\vec\tau\vec\pi)p = p\pi^0+\sqrt{2}n\pi^+
	\,,\end{equation}
where
	\begin{equation}
n=\left( \begin{array}{c} 0 \\ 1 \end{array} \right)
	\,,\end{equation}
($\tau_0 n=-n$) is the isostate of the neutron. That is, the physical
proton can be found in a superposition of a bare-proton plus $\pi^0$ state
and a bare-neutron plus $\pi^+$ state.

The Schr\"odinger equation~(\ref{sch-3}) can be solved by the gaussian
method~\cite{suzuki-varga,mitroy,fedorov-analytic} where the
integro-differential boundary
condition problem is reformulated as a
generalized eigenvalue problem.
The radial function $\varphi(r)$ is represented as a linear
combination of gaussians (which satisfy the boundary
conditions~(\ref{bcond})),
	\begin{equation}
\varphi(r)= \sum_{n=1}^K c_n e^{-\alpha_n r^2}
	\,,\end{equation}
where $K$ is the number of gaussians in the expansion\footnote{For this
one-dimensional problem the ground state wave-function needs around
four to five gaussians to converge.}.  The $\psi_{\N\pi}$ is then given as
	\begin{equation}
\psi_{\N\pi} =\sum_{n=1}^K c_n \psi_{\N\pi}^{\alpha_n}
	\,.\end{equation}
where
	\begin{equation}
\psi_{\N\pi}^{\alpha} = (\vec\tau\vec\pi)(\vec\sigma\vec r)
\frac{p\up}{\sqrt{V}}e^{-\alpha r^2}
	\,.\end{equation}
The function $\psi_\p$ in the center-of-mass frame of the proton  is
simply a constant,
	\begin{equation}
\psi_\p=c_0\frac{p\up}{\sqrt{V}}
	\,.\end{equation}
The Schr\"odinger equation~(\ref{sch-3}) with the boundary
conditions~(\ref{bcond}) now turns into a generalized matrix eigenvalue
problem,
	\begin{equation}
\mathcal H c = \tilde E\mathcal N c \,,
	\end{equation}
where
	\begin{equation}
c = \left( \begin{array}{c} c_0 \\ c_1 \\ \vdots \\ c_K \end{array}
\right)
	\end{equation}
is the vector of coefficients to be found by solving the eigensystem,
and where the matrix elements of the (symmetric positive definite)
overlap matrix ${\mathcal N}$ are given as (cf.~\cite{fedorov-analytic})
	\begin{eqnarray}
{\mathcal N}_{00} &=& \langle \psi_\p |\psi_\p \rangle = 1 \;,\\
{\mathcal N}_{0i} &=& 0 \;,\\
{\mathcal N}_{ij} &=&
\langle \psi_{\N\pi}^{\alpha_i}|\varphi_{\N\pi}^{\alpha_j}
\rangle =3 \frac{3}{2}\frac{1}{\alpha_i+\alpha_j}
\left(\frac{\pi}{\alpha_i+\alpha_j}\right)^{3/2} \;,
	\end{eqnarray}
where $i,j = 1,\dots,K$.

Assuming for simplicity that the formfactor $f(r)$ is
a gaussian,
	\begin{equation}
f(r)=\mathcal A e^{-\kappa r^2}
	\,,\end{equation}
the matrix elements of the (symmetric) Hamiltonian matrix
${\mathcal H}$ are given as
	\begin{eqnarray}
{\mathcal H}_{00} &=& \langle \psi_\p |K_\p|\psi_\p \rangle =0 \;,\\
{\mathcal H}_{0i} &=&\langle\varphi_{\N\pi}^{\alpha_i}| W |\varphi_p\rangle
=3\mathcal A
\frac32
\frac{1}{\alpha_i+\kappa}
\left(\frac{\pi}{\alpha_i+\kappa}\right)^{3/2} \;,\\
{\mathcal H}_{ij} &=& \langle \varphi_{\N\pi}^{\alpha_i}|
K_\N+K_\pi+m_\pi
|\varphi_{\N\pi}^{\alpha_j} \rangle 
=
3 \frac{\hbar^2}{2m_{\N\pi}}
15\frac{\alpha_i\alpha_j}{(\alpha_i+\alpha_j)^2}
\left(\frac{\pi}{\alpha'+\alpha}\right)^{3/2}
+m_\pi{\mathcal N}_{ij}
	 \;,\end{eqnarray}
where $i,j = 1,\dots,K$.

The generalized eigenvalue problem can be easily solved by standard
methods~\cite{gsl}.

The range parameters $\alpha_k$ of the gaussians are often chosen
stochastically using random~\cite{suzuki-varga} or
quasi-random~\cite{fedorov-quasi} sequences.  However for the current
one-dimensional problem we simply perform global optimization in
the space of $\alpha_k$ using the Melder-Need downhill simplex
method~\cite{gsl,nelder-mead}.

In the following we shall parametrize $\kappa$ and $\mathcal A$ as
   \begin{equation}
\kappa \doteq \frac1{b_w^{2}}\;,\, \mathcal A \doteq \frac{S_w}{b_w}
   \,,\end{equation}
where $b_w$~[fm] and $S_w$~[MeV] are the parameters of the model.

\section{Pion photoproduction off proton}

A photon with the energy in excess of the pion mass (plus recoil) can
knock off the virtual pion from the dressed proton in what is called the
pion photoproduction reaction.  Since the wavelength of a photon with the
energy $\hbar\omega~\gs~\m_\pi$ is on the order of the size of
the dressed proton we have to calculate the cross-section without
the use of the usual multipole expansion of the electromagnetic
operator~(cf.~\cite{matsumoto}).

The interaction between the electromagnetic field with potential $\vec A$
in the radiation gauge and a
particle with charge $e$, mass $m_e$, position $\vec r_e$ and momentum
$\vec p_e$ is given as (neglecting the term of the second order in the
field and the contribution from the magnetic moment of the particle)
	\begin{equation}
V_{\vec A} = \frac{-e}{m_e c}\vec A(\vec r_e,t)\vec p_e
	\;.\end{equation}

The free electromagnetic field can be represented as a superposition of
normal modes.  With plane waves as normal modes the
electromagnetic potential $\vec A(\vec r,t)$ of the free field is given
(in the radiation gauge) as
	\begin{equation}
\vec A(\vec r,t) = \sum_{\vec k,\lambda} \sqrt{\frac{2\pi \hbar
c^2}{\omega_{k}V}} \left( a_{\vec k\lambda}\vec e_{\vec k\lambda}
e^{+i(\vec k\vec r-\omega_k t)} + a_{\vec k\lambda}^\dagger \vec e_{\vec
k\lambda}^{~*} e^{-i(\vec k\vec r-\omega_k t)} \right) \;,
	\end{equation}
where $\vec{k}$, $\omega_k=kc$,
$\lambda$, $\vec{e}_{\vec k\lambda}$ are the wavenumber, frequency,
polarization index, and polarization vector of the normal mode;
$a_{\vec k\lambda}$, $a_{\vec k\lambda}^\dagger$ are the (bosonic)
annihilation/generation operators of a photon with the corresponding
quantum numbers; and where the normal modes are normalized to one photon
in the volume $V$.

In the case of the
neutral pion photoproduction off a proton
it is the proton that interacts with the electromagnetic field. The
corresponding interaction operator is given as
	\begin{equation}
V_{\vec A}
= \frac{-e}{m_p c}\vec A(\vec r_p,t)\vec p_p
= \frac{-e}{m_p c}\vec A\left(\vec R-\frac{m_\pi}{m_p+m_\pi}\vec r,t\right)
\left( \frac{m_p}{m_p+m_\pi}\vec P - \vec p \right) \,,
	\end{equation}
where $e$ is the charge of the proton, $\vec r_p$ is the coordinate of
the proton, $\vec p_p$ is the proton momentum operator,
$\vec r=\vec r_\pi-\vec r_p$ is the relative pion-proton coordinate,
$\vec R=(m_p\vec r_p+m_\pi\vec r_\pi)/(m_p+m_\pi)$ is the coordinate
of the center of mass of the pion-proton system,
$\vec p$ is the
relative pion-proton momentum operator, and $\vec P$ is the total
pion-proton momentum operator.

We assume that when driven away from its virtual neutral pion by the
electromagnetic field the bare proton gets dressed immediately by a new
virtual pion. We shall therefore use the physical mass, $m_p$, of the
proton in the corresponding formulae.

The initial state of the pion photoproduction reaction consists of a
dressed proton and a
plane-wave photon with wavenumber $\vec k$ and polarisation $\lambda$
in the state $a_{\vec{k}\lambda}^\dagger|0\rangle$, where $|0\rangle$
is the electromagnetic vacuum.  In the final state there is a physical
proton plus a neutral pion (in a relative plane-wave motion)
and no more photons, that is, the electromagnetic
vacuum $\langle 0|$.
The electromagnetic part of the matrix element of this transition is
given as
	\begin{equation}
\left\langle 0\left|\frac{-e}{m_p c}
\vec A\left(\vec R-\frac{m_\pi}{m_p+m_\pi}\vec r,t\right)
a_{\ve{k}\lambda}^\dagger \right|0 \right\rangle
=
\frac{-e}{m_p}\sqrt{\frac{2\pi \hbar }{\omega_{k}V}}
\vec e_{\vec k\lambda}
e^{i\vec k(\vec R-\frac{m_\pi}{m_p+m_\pi}\vec r)-i\omega_k t} \;.
	\end{equation}

The probability per unit time of the transition from the initial to the
final state is then given by the Fermi's golden rule,
	\begin{equation}
dw_{fi}=\frac{2\pi}{\hbar}|M_{fi}|^2d\rho_{f}\;,
	\end{equation}
where $d\rho_{f}$ is the density of the final states and where
the transition matrix element is given as (in the lab frame where
$\vec P=0$)
	\begin{equation}
M_{fi} = 
\frac{+e}{m_p}\sqrt{\frac{2\pi \hbar }{\omega_{k}V}}
\left\langle p \xi\pi^0\frac{e^{i\vec q\vec r}}{\sqrt{V}}
\frac{e^{i\vec Q\vec R}}{\sqrt{V}}\right|
e^{i\vec k(\vec R-\frac{m_\pi}{m_p+m_\pi}\vec r)}
(\vec e_{\vec k\lambda}\vec p)
\left|(\vec\tau\vec\pi)(\vec\sigma\vec r)\varphi(r)\frac{p\up}{\sqrt{V}}
\right\rangle \;,
	\end{equation}
where $\xi$ is the
spin-state of the final proton (can be either $\up$ or $\down$),
$\hbar\vec q$ is the pion-proton relative momentum in the final state,
and $\vec Q=\vec k$ is the recoil.

Integration over $d^3R$ gives $V$,
the isospin factor gives
	\begin{equation}
\langle p\pi^0|\vec\tau\vec\pi|p\rangle = 1 \;,
	\end{equation}
and,
recalling that $\vec p = -i\hbar\frac{\partial}{\partial\vec r}$,
the matrix element becomes
	\begin{equation}
M_{fi} =
\frac{-ie\hbar}{m_p V}\sqrt{\frac{2\pi \hbar }{\omega_{k}}}
\left\langle\xi \left|
\big\langle e^{i\vec s\vec r}|
\left(\vec e_{\vec k\lambda}\frac{\partial}{\partial\vec r}\right)
(\vec\sigma\vec r)|\varphi(r)\big\rangle\right|\up\right\rangle
	\;,\end{equation}
where $\vec s\doteq \vec q+\frac{m_\pi}{m_p+m_\pi}\vec k$.

In the inner matrix element the integral over angles can be evaluated
analytically: integrating by parts once and then using the following
formulae,
	\begin{eqnarray}
e^{-i\vec s\vec r}&=&
\sum_l (2l+1) (-i)^l j_l(sr)P_l\left(\frac{\vec s\vec r}{sr}\right) \,,\\
\int d\Omega~r_i r_j &=& \frac{4\pi}{3}r^2\delta_{ij}  \,,
	\end{eqnarray}
(where $P_l$ is the Legendre polynomial and $j_l$ is the spherical
Bessel function) gives
	\begin{eqnarray}
\left\langle e^{i\vec s\vec r}\left|
\left(\vec e\frac{\partial}{\partial\vec r}\right)
(\vec\sigma\vec r)\right|\varphi(r)\right\rangle
&=&+i(\vec e\vec s) \int d^3r e^{-i\vec s\vec r}
(\vec\sigma\vec r) \varphi(r) \\
&=&-i(\vec e\vec s)
\int d^3r 3i j_1(sr)
\frac{(\vec s\vec r)}{sr}
(\vec\sigma\vec r)
\varphi(r) \\
&=&(\vec e\vec s)(\vec\sigma\vec s)\frac{4\pi}{s}
\int_0^\infty r^3 dr j_1(sr)\varphi(r)
	\,.\end{eqnarray}
Thus the inner matrix element is given as
	\begin{equation}
\big\langle e^{i\vec s\vec r}\big|
\left(\vec e\frac{\partial}{\partial\vec r}\right)
(\vec\sigma \vec r)\big|\varphi(r)\big\rangle
=(\vec e \vec s)(\vec\sigma \vec s)F(s)
	\,.\end{equation}
where the factor $F(s)$
is determined by the spherical Bessel transform of $\varphi(r)$,
	\begin{equation}
F(s)=\frac{4\pi}{s}\int_0^\infty r^3 dr j_1(sr)\varphi(r)
	\,.\end{equation}

If $\varphi(r)$ is represented as a linear combination of gaussians,
	\begin{equation}
\varphi(r)=\sum_n c_n e^{-\alpha_n r^2} \,,
	\end{equation}
the radial integral in $F(s)$ can be calculated analytically
either using the formula
	\begin{equation}
\frac{4\pi}{s}\int_0^\infty r^3 dr j_1(sr)e^{-\alpha r^2}
=\frac{1}{2\alpha}e^{-\frac1{4\alpha}s^2}\left(\frac{\pi}{\alpha}\right)^{3/2}
	\,,\end{equation}
which gives
	\begin{equation}
F(s)= \sum_n c_n
\frac{4\pi}{s}\int_0^\infty r^3 dr j_1(sr)e^{-\alpha_n r^2}
= \sum_n c_n 
\frac{1}{2\alpha_n}e^{-\frac1{4\alpha_n}s^2}\left(\frac{\pi}{\alpha_n}\right)^{3/2}
	\,,\end{equation}
or like this\footnote{
First the matrix element with the operator $r_ir_j$ is calculated as
	\begin{eqnarray}
\langle e^{i\vec s\vec r}| r_i r_j|e^{-\alpha r^2}\rangle
=\frac{-\partial^2}{\partial s_i\partial s_j}
\langle e^{i\vec s\vec r}|e^{-\alpha r^2}\rangle
=\frac{-\partial^2}{\partial s_i\partial s_j}
e^{-\frac{1}{4\alpha}s^2}\left(\frac{\pi}{\alpha}\right)^{3/2}
=\left(\frac{1}{2\alpha}\delta_{ij}-\frac{1}{4\alpha^2}s_is_j\right)
e^{-\frac{1}{4\alpha}s^2}\left(\frac{\pi}{\alpha}\right)^{3/2}
	\,.\end{eqnarray}
And now the sought matrix element evaluates as
	\begin{eqnarray}
\langle e^{i\vec s\vec r}|
(\vec e\frac{\partial}{\partial\vec r})
(\vec \sigma\vec r)
|e^{-\alpha r^2}\rangle
=(\vec e\vec \sigma) \langle e^{i\vec s\vec r}| e^{-\alpha r^2}\rangle
-2\alpha\langle e^{i\vec s\vec r}
|(\vec e\vec r)(\vec \sigma\vec r)|e^{-\alpha r^2}\rangle
\nonumber\\ =
(\vec e\vec \sigma)e^{-\frac{1}{4\alpha}s^2}
\left(\frac{\pi}{\alpha}\right)^{3/2}
-2\alpha\left(
\frac{1}{2\alpha}(\vec e\vec \sigma)
-\frac{1}{4\alpha^2}(\vec e\vec s)(\vec \sigma\vec s)
\right)
e^{-\frac{1}{4\alpha}s^2}\left(\frac{\pi}{\alpha}\right)^{3/2}
\nonumber\\ =
\frac{1}{2\alpha}(\vec e\vec s)(\vec \sigma\vec s)
e^{-\frac{1}{4\alpha}s^2}\left(\frac{\pi}{\alpha}\right)^{3/2}
	\end{eqnarray}
}.

The total matrix element is now given as
	\begin{equation}
M_{fi} = 
\frac{-ie\hbar}{m_p V}\sqrt{\frac{2\pi \hbar }{\omega_{k}}}
(\vec e_{\vec k\lambda}\vec s)
\langle\xi |
\vec \sigma\vec s
|\up\rangle F(s)
	\;.\end{equation}
The transition probability is determined by the absolute square of the
matrix element,
	\begin{equation}
|M_{fi}|^2 =
\frac{2\pi}{V^2}
\frac{e^2\hbar^3}{m_p^2\omega_{k}}
\big|(\vec e_{\vec k\lambda}\vec s)\big|^2
\big|\langle\xi |
\vec \sigma\vec s
|\up\rangle\big|^2
F^2(s)
	\;.\end{equation}
Assuming that the beam and the target are unpolarized we need to average
the transition probability
over the polarizations of the photon (and the target) in the initial
state and sum over the
spin states of the proton in the final state. Summation over the photon
polarizations gives
	\begin{equation}
\sum_\lambda (\vec e_{\vec k\lambda}^{~\star}\vec s) (\vec e_{\vec
k\lambda}\vec s) =s^2-\frac{(\vec k\vec s)^2}{k^2} =q^2-\frac{(\vec
k\vec q)^2}{k^2}=q^2\sin^2(\theta_q)
	\,,\end{equation}
where $\cos(\theta_q)=(\vec q~\vec k)/(qk)$.

Summation over the spin states of the final proton gives,\footnote{
	\begin{equation}
\langle\up | \vec \sigma\vec s |\up\rangle = s_z
\,,\,
\langle\down | \vec \sigma\vec s |\up\rangle = s_x+is_y
\,,\,
s_z^2+(s_x-is_y)(s_x+is_y)=s^2
	\,.\end{equation}
}
   \begin{equation}
\sum_\xi \left|\langle\xi|~\vec\sigma\vec s~|\up\rangle\right|^2 = s^2
   \,.\end{equation}
We do not really need to average over the spin states of the final proton
as the two states give identical contribution.
The averaged matrix element is given as
	\begin{equation}
\frac12 \sum_\lambda \sum_\xi
|M_{fi}|^2 =
\frac{\pi}{V^2}
\frac{e^2\hbar^3}{m_p^2\omega_{k}}
q^2\sin^2(\theta_q)
s^2 F^2(s)
	\;.\end{equation}

The number of final states around $\vec q$ is given as
	\begin{equation}
dn_{\vec q}
= \frac{Vd^3q}{(2\pi)^3}
= \frac{V d\Omega_q}{(2\pi)^3}q^2 dq 
= \frac{V d\Omega_q}{(2\pi)^3}\frac12 q dq^2 \;,
	\end{equation}
which gives the density of the final states as
	\begin{equation}
d\rho_f = \frac{dn_{\vec q}}{dE_q}
= \frac{V d\Omega_q}{(2\pi)^3}q \frac12 \frac{dq^2}{dE_q}
	\,,\end{equation}
where $E_q$ is the energy of the relative motion of the particles in
the final state.
The averaged transition probability is then given as
	\begin{equation}
dw_{\mathrm fi}=\frac{2\pi}{\hbar}
\frac12 \sum_\lambda \sum_\xi
|M_{\mathrm fi}|^2 d\rho_{\mathrm f} =
\frac{2\pi}{\hbar}
\frac{\pi}{V^2}
\frac{e^2\hbar^3}{m_p^2\omega}
q^2\sin^2(\theta_q)
s^2 F^2(s)
\frac{V d\Omega_q}{(2\pi)^3}q \frac12 \frac{dq^2}{dE_q}
	\,.\end{equation}

Finally, division by the flux density of the incoming photons, $c/V$,
gives the differential cross-section for the
photoproduction of neutral pion off protons,
	\begin{equation}
\frac{d\sigma(E_q,\theta_q)}{d\Omega_q} =
\frac{e^2}{8\pi}
\frac{1}{\m_p^2}
\frac{q^3}{k}
\frac{d(\hbar qc)^2}{dE_q}
\sin^2(\theta_q)
s^2 F^2(s) \,.
	\end{equation}

The energy $E_q$ of the relative pion-proton motion in the final states
is given by the (relativistic, for generality) energy conservation
relation,
	\begin{equation}
E_q = \hbar\omega + \m_p - \sqrt{(\m_p+\m_\pi)^2+\hbar^2\omega^2}
	\,.\end{equation}
The absolute value of the relative wave-number $\vec q$ can then be
calculated using the energy equation in the final state,
	\begin{equation}
E_q=\sqrt{\m_p^2+(\hbar qc)^2}+\sqrt{\m_\pi^2+(\hbar qc)^2}
-\m_p-\m_\pi
	\,,\end{equation}
which gives
	\begin{equation}
(\hbar qc)^2 =
\frac{E_q(E_q+2\m_p)(E_q+2\m_\pi)(E_q+2\m_p+2\m_\pi)}{4(E_q+\m_p+\m_\pi)^2}
	\,,\end{equation}
and
	\begin{equation}
\frac{d(\hbar qc)^2}{dE_q}=
\frac{
(E_q^2+2E_q\m_p +2\m_p^2+2 E_q\m_\pi + 2\m_p\m_\pi)
(E_q^2+2E_q\m_p +2\m_{\pi}^2+2 E_q\m_\pi + 2\m_p\m_\pi)
}{2(E_q+\m_p+\m_\pi)^3}
	\,.\end{equation}

The total cross-section $\sigma$ is given as the integral over $\theta_q$,
   \begin{equation}
\sigma = \int_0^\pi 2\pi \sin(\theta_q)d\theta_q \frac{d\sigma}{d\Omega_q}
   \,.\end{equation}

\section{Results}

\begin{figure}
\centerline{\includegraphics{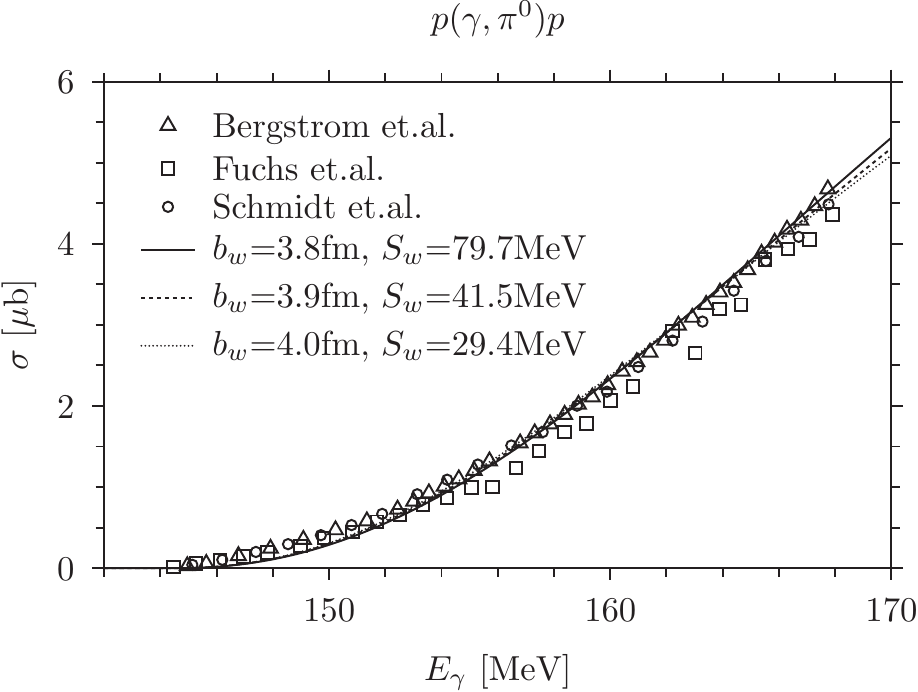}}
\caption{Total cross section, $\sigma$, for the neutral pion
photoproduction off
a proton as function of photon energy, $E_\gamma$, calculated with
several sets
of model parameters and shown together with the experimental data
from~\cite{bergstrom,fuchs,schmidt}.  The errorbars are about the size
of the symbols.}
   \label{fig-zero} \end{figure}

The calculated total cross section for neutral pion photoproduction off a
proton as function of the photon energy is shown on Figure~\ref{fig-zero}
together with the recent experimental data.  The model is apparently
able to quantitatively describe the experiment.  However the range of
the parameters, where the model agrees with the experiment, is relatively
broad: basically for any $b_w$ in the range from 3.7fm to 4.0fm one can
find the corresponding strength $S_w$ that would fit the data (perhaps
with $b_w=$3.9fm being the favourite).

For the set of parameters $b_w$=3.9fm, $S_w$=41.5MeV the relative weight
of the $\pi^0$ component in the wave-function of the dressed proton is
about 25\% and the virtual pions contribution to the mass of the dressed
proton is about -586MeV.  All these numbers might seem somewhat excessive
and might indicate that the two-pion effects are not negligible or that
our phenomenological coupling operator is not good enough.
Calculations of the deuteron with the found parameters might provide
a better insight into the validity of this model.

\section{Conclusion}
We have applied the nuclear model with explicit mesons to pion
photoproduction off protons.  In this model the nucleons do not interact
directly with each other but rather emit and absorb mesons that are
treated explicitly on equal footing with the nucleons.  The mesons emitted
by a bare nucleon find themselves in a classically forbidden region under
a potential barrier about the size of the mass of the mesons.  The mesons
cannot leave the nucleon and can thus be referred to as virtual.  A bare
nucleon surrounded by virtual mesons is the physical or dressed nucleon.
We have considered the dressing of a proton with pions in the one-pion
approximation and derived the equations for the wave-function of the
virtual pion.

A photon with the energy in excess of the pion mass (plus recoil) can
knock off a virtual pion from the dressed proton.  That would be the
pion photoproduction reaction.  We have calculated the cross section
of the neutral pion photoproduction off protons near threshold and have
compared it with the experimental data.  We have shown that the model
is able to quantitatively describe the experiment for a certain range
of parameters.

In conclusion we have shown that the nuclear model with explicit mesons
is able to reproduce quantitatively the experimental cross section for
neutral pion photoproduction off protons close to threshold, and have
determined the range of the parameters where the model is compatible
with the experimental data.

\section{Acknowledgments}
Numerous discussions with Hans Fynbo and Karsten Riisager are
gratefully acknowledged.

\end{document}